\documentclass[useAMS,usenatbib,referee]{biom}

\usepackage{bm}
\usepackage{amssymb,mathtools}
\usepackage{natbib}
\usepackage[hyphens]{url}
\usepackage[linesnumbered, ruled]{algorithm2e}
\usepackage{hyperref}
\hypersetup{
colorlinks=true,
linkcolor=red,
filecolor=magenta,
citecolor=blue,
urlcolor=blue,
}
\usepackage{cleveref}
\Crefname{assumption}{Assumption}{Assumptions}
\usepackage{setspace}
\usepackage{tikz}
\usepackage{calc} 
\tikzset{>=latex} 

\usepackage{bbm}
\usepackage{textcomp} 

\usepackage{multirow,multicol,makecell,booktabs}

\usepackage[figuresright]{rotating}
\usepackage{xcolor,colortbl}

\definecolor{Gray}{gray}{0.80}
\definecolor{Gray2}{gray}{0.95}
\def\bSig\mathbf{\Sigma}

\title[]{Incorporating External Risk Information with the Cox Model under Population Heterogeneity: Applications to Trans-Ancestry Polygenic Hazard Scores}

\author{Di Wang$^{1}$, Wen Ye$^{1}$, Ji Zhu$^{2}$, Gongjun Xu$^{2}$, Weijing Tang$^{3}$, Matthew Zawistowski$^{1,4}$,
\\ \textbf{Lars G. Fritsche}$^{\textbf{1,4,5}}$,
\textbf{and Kevin He$^{\textbf{1,*}}$}\email{kevinhe@umich.edu} \\ $^{1}$Department of Biostatistics, University of Michigan, Ann Arbor, Michigan, U.S.A. \\ $^{2}$Department of Statistics, University of Michigan, Ann Arbor, Michigan, U.S.A \\ $^{3}$Department of Biostatistics, Harvard University, Boston, Massachusetts, U.S.A.\\ $^{4}$Center for Statistical Genetics, University of Michigan, Ann Arbor, Michigan, U.S.A. \\
$^{5}$Rogel Cancer Center, University of Michigan, Ann Arbor, Michigan, U.S.A.}

\begin{document}





\pagerange{\pageref{firstpage}--\pageref{lastpage}} 




\label{firstpage}

\begin{abstract}
Polygenic hazard score (PHS) models designed for European ancestry (EUR) individuals provide ample information regarding survival risk discrimination. Incorporating such information can improve the performance of risk discrimination in an internal small-sized non-EUR cohort. However, given that external EUR-based model and internal individual-level data come from different populations, ignoring population heterogeneity can introduce substantial bias. In this paper, we develop a Kullback-Leibler-based Cox model (CoxKL) to integrate internal individual-level time-to-event data with external risk scores derived from published prediction models, accounting for population heterogeneity. Partial-likelihood-based KL information is utilized to measure the discrepancy between the external risk information and the internal data. We establish the asymptotic properties of the CoxKL estimator. Simulation studies show that the integration model by the proposed CoxKL method achieves improved estimation efficiency and prediction accuracy. We applied the proposed method to develop a trans-ancestry PHS model for prostate cancer and found that integrating a previously published EUR-based PHS with an internal genotype data of African ancestry (AFR) males yielded considerable improvement on the prostate cancer risk discrimination.

\end{abstract}

%



\maketitle


%

\section{Introduction}
\label{s:intro}

Survival risk discrimination is important for identifying individuals with a high risk of disease progression. For example, polygenic hazard scores (PHS), calculated as the sum of single-nucleotide polymorphisms (SNPs) weighted by the risk allele effect sizes derived from Cox proportional hazard models, are effective tools for personalized disease risk assessment and are widely applied in cancer research \citep{Seibert2018,Zhang2020}. However, the development of PHS has often included only individuals of European ancestry (EUR). A major challenge in developing PHS models for a minority population is that non-EUR individuals are usually less represented in genetic databases \citep{Landry2018},
limiting their utility and exacerbating existing health disparities  \citep{Martin2017}. 

Our research is motivated by the study of prostate cancer, which is the second most prevalent cancer in males across the world \citep{Bray2018}. Recently, a PHS calculated with 46 SNPs (denoted as PHS46) has been developed to predict the age of onset of prostate cancer in EUR individuals \citep{Huynh2021}. However, it is well-known that African ancestry (AFR) individuals have higher prostate cancer incidences and mortality rates than EUR individuals \citep{DeSantis2016}, which may be partially explained by disparities in gene mutations associated with prostate cancer \citep{Huang2017}. Due to these clinical and genetic heterogeneity, directly applying EUR-based PHS has limited utility for non-EUR populations \citep{vilhjalmsson2015modeling}. 
On the other hand, constructing PHS for non-EUR populations suffers from limited sample sizes, high dimensionality, and low signal-to-noise ratios. Thus, it would be desirable to develop a method that integrates risk information from published prediction models upon EUR individuals (e.g. PHS46) with individual-level data collected in non-EUR cohorts (e.g. genotype data of AFR males), automatically measure the heterogeneity across different populations, and efficiently balance the contribution of each information source.

Ideally, the access to individual-level data will enable the pooled analysis and optimize the use of each information source. However, due to restrictions on data privacy, summary-level information is commonly used for data sharing in most applications. In the context of survival analysis, several methods for integrating summary-level external information with an internal study have been proposed. For example, \citet{huang2016} proposed a double empirical likelihood approach based on a homogeneous assumption; i.e. both the covariate distribution and the association between  covariates and outcomes are the same in the internal and external cohorts. However, this strong assumption is rarely met. Ignoring heterogeneity across different studies yields substantially biased parameter estimates and a loss of efficiency \citep{estes17}. 

To relax the homogeneity assumption, 
\citet{chen2021} proposed an adaptive double empirical likelihood estimator by penalizing potential differences on the subgroup survival probabilities. 
\citet{sheng2021:2} constructed a heterogeneity test based on the penalized empirical likelihood and proposed a semiparametric density ratio model. 
Although these methods are useful, they typically require external subgroup survival probabilities, which are not available through EUR-based PHS models. The published PHS contains only covariate coefficients, but with no baseline hazard information. Such risk scores cannot be given any direct probability interpretation. Furthermore, it can be computationally expensive for existing methods to handle large-scale data, especially when the number of predictors largely exceeds the sample size (e.g. non-EUR genotype data), leading to a high demand for new integration methods that are computationally efficient and enable full utilization of high-dimensional information.

Kullback-Leibler (KL) information \citep{KL} has been widely applied in the development of data integration methods for binary and ordinal outcomes under population heterogeneity \citep{Liu2003, Schapire2005, plasso, hector22}. However, these methods apply only to generalized linear models (GLM). For the integration with survival outcomes, \citet{wang20212} extended the GLM-based KL to discrete failure time models and developed an integration procedure to incorporate external survival probabilities across multiple time points with internal time-to-event data. However, it is not clear how to incorporate external survival risk information without probability interpretation. To fill this gap, we propose a KL-based integrated Cox model (CoxKL), which incorporates external risk scores with internal individual-level time-to-event data to improve the estimation and prediction performance in the internal cohort. 
A key ingredient of the proposed method is the partial-likelihood-based KL information, which measures the heterogeneity between the external and the internal populations. We first develop the CoxKL method under low-dimensional settings, and then extend it to CoxKL-LASSO with a $L_{1}$ penalization, which can be further applied to high-dimensional trans-ancestry PHS developments. 

The proposed CoxKL procedure attains the following advantages: (1) it only requires summary-level risk scores to be obtained from external studies, but with no use of any individual-level external data or survival probabilities, and therefore achieves greater flexibility in data sharing and integrative analysis; (2) it accounts for population heterogeneity,
automatically detects the quality of different information sources, and provides an optimal integrated model based on the aggregated information;
and (3) its objective function retains a similar form of the log-partial likelihood of the classical Cox model; thus, it is computationally efficient for high-dimensional problems, which is crucial for the motivating trans-ancestry PHS analysis; and (4) it provides a unified integration framework to incorporate risk information from various types of external prediction models or multiple external models. 

The rest of the paper is organized as follows: In Section~\ref{s:notation}, we introduce the notations and model setup. In Section~\ref{s:method}, we propose the CoxKL method. Asymptotic properties are discussed in Section~\ref{s:property}. Section~\ref{s:simu} conducts simulations to evaluate the proposed method. In Section~\ref{s:real_data}, we applied CoxKL method to develop a trans-ancestry PHS model for prostate cancer in AFR males. Section~\ref{s:dis} concludes with a brief discussion.

\section{Notation and Model Setup}
\label{s:notation}
In Section~\ref{sub:internal}, we introduce the notations of the Cox proportional hazards model for the target internal cohort, which collects individual-level data with time-to-event outcomes and a set of covariates from a limited-size sample (denoted as internal data). In Section~\ref{sub:external}, we describe various types of external risk scores 
that can be integrated by the CoxKL method.

\subsection{Internal Cox proportional hazards model}
\label{sub:internal}
Let $T_i$ denote the event time of interest and $C_i$ be the  censoring time for subject $i$, $i=1,\ldots, n$, where $n$ is the sample size of the internal cohort. Let $\bm{Z}_{i}$ be a $p$-dimensional covariate vector for the $i$-th subject. We assume that, conditional on $\bm{Z}_i$,  $T_i$ is independently censored by  $C_i$. The observed time is denoted by $X_i=\mbox{min}\{T_i, C_i\}$, and the
event indicator is given by $\delta_i = I(T_i \leq C_i)$. 
The internal cohort consists of $n$ independent vectors
$(X_i,\delta_i,\bm{Z}_i)$. 
Consider a hazard function specified by
  \begin{eqnarray}
     \lambda(t|\bm{Z}_i) = \lim_{dt\rightarrow 0} \frac{1}{dt}Pr(t \leq T_i < t+dt |  T_i \geq t,  \bm{Z}_i) = 
\lambda_{0}(t)\exp{(\bm{Z}_i^\top\bm{\beta})}, \nonumber
\end{eqnarray}
where $\lambda_{0}(t)$ is the baseline hazard function,  
and $\bm{\beta}=(\beta_{1},\cdots, \beta_{p})^\top$ is a $p$-dimensional regression parameter. Suppose the internal cohort comprises $K$ unique failure times $t_1< \cdots < t_K$.
Considering a Cox proportional hazards model \citep{cox1972regression}, the corresponding partial likelihood for $\bm{\beta}$ is given by
  \begin{eqnarray}
L(\boldsymbol\beta)=
\prod_{k=1}^K 
\frac{\exp{(\bm{Z}_k^\top\bm{\beta})}}{\sum_{i=1}^n Y_i(t_k)\exp{(\bm{Z}_i^\top\bm{\beta})}},
 \label{eq:partial}
\end{eqnarray}
where $\bm{Z}_k$ is the covariate vector for the subject who experiences event at $t_k$, and $Y_i(t_k)=I(X_i \geq t_k)$ is the at-risk indicator. The log-partial likelihood is given by
\begin{eqnarray}
\ell(\boldsymbol\beta)&=&{}\sum_{k=1}^K  \left [\bm{Z}_k^\top\bm{\beta} -\log \left\{\sum_{i=1}^n Y_i(t_k) \exp{(\bm{Z}_i^\top\bm{\beta})}\right \} \right ].
\label{eq:logpartial}
\end{eqnarray}

\subsection{External risk scores}
\label{sub:external}

Due to restrictions on data privacy, the access to the individual-level genotype data from the external EUR studies is usually limited. Only summary-level information, such as coefficient estimates $\widetilde{\boldsymbol\beta}$, can be obtained from the published PHS model. Specifically, we consider the situation that some external risk scores, $r(\bm{Z}_i, \widetilde{\boldsymbol\beta})$, can be computed based on published EUR-based PHS models, where $\bm{Z}_i$ is the input covariate vector for the $i$-th patient in the internal non-EUR cohort. 

In practice, the variables contained in the internal data are not necessarily the same as those used in the external risk scores. For example, the PHS46 model only utilized 46 SNPs, but the internal genotype data of AFR individuals contains millions of SNPs after imputation. By allowing $\widetilde{\bm{\beta}}$ to be a sparse vector (i.e. some entries of $\widetilde{\bm{\beta}}$ can be zero), the proposed method can 
be applied to the case that the internal data and the external risk score share the same set of covariates, as well as the case that the internal data contains additional covariates. 

In addition to accommodating for differences in the covariate space, another advantage of external risk scores is that it can incorporate different forms of external risk score models, such as the commonly used Cox proportional hazards model, with $r(\bm{Z}_i, \widetilde{\boldsymbol\beta})=\bm{Z}_i^\top \widetilde{\boldsymbol\beta}$, which assumes a linear and proportional effect. Moreover, $r(\bm{Z}_i, \widetilde{\boldsymbol\beta})$ can also be obtained from machine learning algorithms
\citep{Gao2023}, as long as the external model provides a predicted risk score for each subject of the internal population.

\section{Proposed Methods}
\label{s:method}

We propose using KL information to evaluate the disparity between the internal data and external risk scores, and consequently integrate the optimally weighted external information with the internal data to obtain the best integrated model for the target internal population. We present CoxKL under low-dimensional settings in Section~\ref{sub:CoxKL}, and then present CoxKL-LASSO, which extends CoxKL to high-dimensional trans-ancestry PHS problems, in Section~\ref{sub:CoxKL-LASSO}.

\subsection{CoxKL: KL-based integrated Cox model}
\label{sub:CoxKL}

Consider the internal Cox model, its partial likelihood in \eqref{eq:partial} arises as a product of conditional probability statements \citep{cox1975, Kalbfleisch2002}.
Specifically, let $A_k$ specify the event that individual $k$ fails in $[t_k, t_k+dt_k)$, and let $B_k$ specify all the censoring and failure information up to time $t_k^-$ as well as the information that one failure occurs in $[t_k, t_k+dt_k)$. Then $A_1|B_1, \dots, A_K|B_K$ is a sequence of conditional experiments, where the $\sigma$-field generated by $B_{k}$ is contained by that generated by $B_{k+1}$. The density function of $a_k$ conditional on $B_k=b_k$ is
\begin{eqnarray}
f(a_k|b_k; \boldsymbol\beta) &=&{} \frac{\lambda(t_k|\bm{Z}_k)dt_k}{\sum_{i=1}^n Y_i(t_k)\lambda(t_k|\bm{Z}_i)dt_k}=
 \frac{\exp(\bm{Z}_k^\top\bm{\beta})}{\sum_{i=1}^n Y_i(t_k) \exp(\bm{Z}_i^\top\bm{\beta})},
\label{eq:f2}
\end{eqnarray}
where the second equality is obtained by canceling  $\lambda_0(t_k)dt_k$ in the numerator and the denominator. 
 To extract information from external risk scores, we adopt a similar conditional density
\begin{equation}
 f(a_k|b_k;  \widetilde{\boldsymbol\beta})=\frac{\exp{\{r(\bm{Z}_k,  \widetilde{\boldsymbol\beta})\}}}{\sum_{i=1}^n Y_i(t_k) \exp{\{r(\bm{Z}_i,  \widetilde{\boldsymbol\beta})\}}}.
\end{equation}
The partial likelihood-based KL information between conditional densities corresponding to the external risk scores and the internal Cox model, contained in $a_k|b_k$, is given by
\begin{equation}
d_{KL}(\widetilde{\boldsymbol\beta} \parallel \bm{\beta} ; t_k) = \mbox{E} \left[\log\left\{\frac{f(a_k|b_k; \widetilde{\boldsymbol\beta})}{ f(a_k|b_k;\bm{\beta})}\right\}\bigg|b_k; \widetilde{\boldsymbol\beta}\right],
\label{eq:KL}
\end{equation}
where the expectation $\mbox{E}$ is with respect to the conditional density $f(a_k|b_k; \widetilde{\boldsymbol\beta})$. The accumulated KL information in the sequence of conditional experiments $a_1|b_1,  \cdots, a_K|b_K$ is then defined as $D_{KL}(\widetilde{\boldsymbol\beta} \parallel \bm{\beta})=\sum_{k=1}^K d_{KL}(\widetilde{\boldsymbol\beta} \parallel \bm{\beta} ;t_k)$, 
which measures the discrepancy between the external risk scores and the internal Cox model for the target cohort. 

To integrate the external risk scores with the internal data, we construct a penalized log-partial likelihood, which combines the log-partial likelihood from the internal Cox model and the accumulated KL information $D_{KL}(\widetilde{\boldsymbol\beta}\parallel \bm{\beta})$, 
\begin{align}
\ell_{\eta}(\bm{\beta})
&= \ell(\bm{\beta})-\eta D_{KL}(\widetilde{\boldsymbol\beta} \parallel \bm{\beta}),
\label{eq:penalized-loglik}
\end{align}
where
$\eta$ is a tuning parameter weighing the relative importance of external and internal data sources. In the extreme case of $\eta = 0$, the penalized log-partial likelihood $\ell_{\eta}(\bm{\beta})$ is reduced to the log-partial likelihood of the internal Cox model $\ell(\bm{\beta})$. 

\noindent
{\textit {Remark 1}}: The goal of the penalized objective function is to balance the trade-off between the external risk scores and the internal data; i.e. maximize the log-partial likelihood of the internal data, while at the same time keeping the KL information small. When internal and external information are homogeneous, it's intuitive that the CoxKL estimator attains better estimation efficiency and higher discrimination power. In contrast, when internal and external information are heterogeneous,  external information introduces bias to the integrated model. However, due to the bias and variance trade-off, it can be shown that there exists a range of $\eta$ within which the CoxKL method improves the estimation efficiency; further details are provided in Section~\ref{s:property}.

\subsection{Estimation}
\label{sub:Estimation}

Proposition 1 below shows that the objective function $\ell_{\eta}(\bm{\beta})$ retains a similar form as the log-partial likelihood of a classical Cox model. Hence, the integrated model can be solved by applying the standard optimization algorithms for the classical Cox models, which ensures the computational efficiency of the proposed method. A proof of Proposition 1 is provided in Supporting Information. 
 
 \noindent
\textbf{Proposition 1}
{\it The accumulated KL information is given by
\begin{eqnarray}
\nonumber
D_{KL}(\widetilde{\boldsymbol\beta} \parallel \bm{\beta})
&\propto&{}-\sum_{k=1}^K \left [ \widetilde{\bm{Z}}_k^\top\bm{\beta} - \log \left\{\sum_{i=1}^n Y_i(t_k) \exp
(\bm{Z}_i^\top\bm{\beta})\right \} \right ],
\end{eqnarray}
where 
\begin{eqnarray}
\nonumber
\widetilde{\bm{Z}}_k=\frac{\sum_{i=1}^n Y_i(t_k) \exp{\{r(\bm{Z}_i, \widetilde{\boldsymbol\beta})\}}\bm{Z}_i}{\sum_{i'=1}^n Y_{i'}(t_k) \exp{\{r(\bm{Z}_{i'}, \widetilde{\boldsymbol\beta})\}}}
\end{eqnarray}
is a weighted average of the covariates for subjects at risk 
at time $t_k$, with weights defined upon external risk scores. Furthermore, the objective function $\ell_{\eta}(\bm{\beta})$ in \eqref{eq:penalized-loglik} is equivalent to
\begin{eqnarray}
\ell_{\eta}(\bm{\beta})
&\propto&{} \sum_{k=1}^K \left [\left\{\frac{\bm{Z}_k+\eta\widetilde{\bm{Z}}_k}{1+\eta}\right\} ^{\top}\boldsymbol\beta - \log \left\{\sum_{i=1}^n Y_i(t_k) \exp(\bm{Z}_i^\top \boldsymbol\beta)\right \} \right].
\label{eq:obj_eta}
\end{eqnarray}
}

\subsection{CoxKL-LASSO: Extending CoxKL with $L_{1}$ penalization}
\label{sub:CoxKL-LASSO}
A key challenge for applying existing integration methods to develop trans-ancestry PHS models is the high-dimension of the genotype data. Fortunately,
the concise form of CoxKL objective function $\ell_\eta(\bm{\beta})$ in \eqref{eq:obj_eta} facilitates its extension to high-dimensional problems, such as the LASSO penalization \citep{Tibshirani1996} that simultaneously achieves variable selection and parameter estimation. 

Specifically, the objective function of CoxKL method with a $L_1$ penalty (denoted as CoxKL-LASSO) is given by
\begin{eqnarray}
        \ell_{\eta, \lambda}(\bm{\beta}) 
        &=&{} \ell_{\eta}(\bm{\beta}) -\lambda \|\boldsymbol\beta\|_1,
    \label{eq:CoxKL-LASSO}
\end{eqnarray}
where $\|\boldsymbol\beta\|_1=\sum_{j=1}^p|\beta_j|$ is the $L_1$ norm of $\boldsymbol\beta$, and
$\lambda$ is a tuning parameter that determines the amount of penalization.
Since the penalized objective function in \eqref{eq:CoxKL-LASSO} has a similar form as the classical Cox Lasso \citep{Tibshirani1997CoxLasso}, 
the parameter estimation can be easily obtained by the commonly used coordinate descent-based procedures \citep{Simon2011Coxlasso}.
Therefore, the proposed CoxKL method is computationally efficient for high-dimensional problems such as developing trans-ancestry PHS models for non-EUR populations.

\noindent
{\textit {Remark 2}}:
LASSO penalization has been widely utilized in polygenic risk scores (PRS) for quantitative and binary traits \citep{Austin2013, Mak2017, PANPRS}. However, we could not directly apply these PRS constructing methods for developing PHS models, since the time-to-event data requires unique approaches for handling censored outcomes. We note that, in addition to trans-ancetry PHS models, the CoxKL-LASSO can also be applied to develop classical PHS (e.g. EUR-based PHS) by setting $\eta=0$. 

\subsection{Tuning parameter selection}
The optimal tuning parameter $\eta$ and $\lambda$ can be determined by conventional cross-validation approaches,  such as the V\&VH approach \citep{Verweij1993}, which computes the difference between the log-partial likelihood evaluated on the whole set and  the training set as a measure of the model performance. This approach ensures the validity and stability of the calculation of log-partial likelihood, and it has been widely used for penalized Cox proportional hazards models \citep{friedman2010,simon2011,dai2019}.

\section{Theoretical Properties}
\label{s:property}
 
In this section, we study the asymptotic distribution of the CoxKL estimator under the low-dimensional setting (i.e. $p$ is fixed and $p \ll n$) and show that the proposed method improves the asymptotic mean squared error (aMSE) given a certain range of $\eta$.

We first provide some prerequisite notations. For a given $\eta$, let $\hat{\bm{\beta}}(\eta)$ be the solution obtained by maximizing the penalized objective function in \eqref{eq:obj_eta}, and let $\bm{\beta}^{*}(\eta)$ be the maximum of the expectation of the penalized objective function $\mathbb{E}\{\ell_{\eta}(\bm{\beta}))\}$. Denote $\hat{\bm{\beta}}$ as the maximum log-partial likelihood estimator of the internal model in \eqref{eq:logpartial}. The true value of the regression parameter for the Cox model using the internal cohort only is denoted by $\bm{\beta}_0$. Define $\bm{\Delta}=\bm{\beta}_0-\widetilde{\bm{\beta}}$, where $\widetilde{{\bm{\beta}}}$ are the estimated coefficients from the external model. For the ease of notations, we assume that the external risk scores can be constructed as $\bm{Z}_i^\top \widetilde{\boldsymbol\beta}$. 
Denote 
\begin{align*}
S^{(0)}(t; \bm{\beta})=&{}\frac{1}{n}\sum_{i=1}^n Y_{i}(t)\exp(\bm{Z}_{i}^\top \bm{\beta}),\\
S^{(1)}(t; \bm{\beta})=&{} \frac{\partial S^{(0)}(t; \bm{\beta})}{\partial \bm{\beta}}=\frac{1}{n}\sum_{i=1}^n Y_{i}(t) \bm{Z}_{i}\exp(\bm{Z}_{i}^\top \bm{\beta}),\\
S^{(2)}(t; \bm{\beta})=&{}\frac{\partial^2 S^{(0)}(t; \bm{\beta})}{\partial \bm{\beta} \partial \bm{\beta}^\top}=\frac{1}{n}\sum_{i=1}^n Y_{i}(t) \bm{Z}_{i}^{\otimes 2}\exp(\bm{Z}_{i}^\top \bm{\beta}),
\end{align*}
 where $\bm{b}^{\otimes2}=\bm{b}\bm{b}^\top$. Then the score function of $\ell_{\eta}(\bm{\beta})$ in \eqref{eq:obj_eta}
 is given by
 \begin{align*}
 \bm{U}_{\eta}(\bm{\beta})=\frac{\partial \ell_{\eta}(\bm{\beta})}{\partial \bm{\beta}}=\frac{1}{n} \sum_{k=1}^K \left \{\frac{\bm{Z}_k+\eta \widetilde{\bm{Z}}_k}{1+\eta} -\frac{S^{(1)}(t_k; \bm{\beta})}{S^{(0)}(t_k; \bm{\beta})} \right \}.
 \end{align*}
The corresponding information matrix is 
 \begin{align*}
 \bm{H}_{\eta}(\bm{\beta})=-\frac{\partial^2 \ell_\eta(\bm{\beta})}{\partial \bm{\beta} \partial \bm{\beta}^\top}= \frac{1}{n} \sum_{k=1}^K \left [\frac{S^{(2)}(t_k; \bm{\beta})}{S^{(0)}(t_k; \bm{\beta})}- \left \{\frac{S^{(1)}(t_k; \bm{\beta})}{S^{(0)}(t_k; \bm{\beta})}\right \}^{\otimes 2}\right ].
  \end{align*}
To derive the theoretical properties of the CoxKL method under the low-dimensional setting, we impose the following
regularity conditions for the internal cohort:
\begin{itemize}
    \item[(C1).] 
$(X_i,\delta_i,\bm{Z}_i)$, $i=1, \ldots, n$, are
 independent and identically distributed random vectors.
    \item[(C2).] $P(X_i \geq \tau)=0$, where $\tau$ is a pre-specified time point.
     \item[(C3).] $Z_{ij}<\kappa$ for $i=1, \ldots, n$, and $j=1, \dots, p$, where $\kappa$ is a constant and $Z_{ij}$ is the $j$-th component of $\bm{Z}_i$.
   \item[(C4).] $\int_0^{\tau}\lambda_0(t) < \infty$.
    \item[(C5).] Continuity of $s^{(d)}(t; \bm{\beta})$, $d=0,1,2$, for any $\bm{\beta}$ between $\bm{\beta}_0$ and $\widetilde{\bm{\beta}}$,
   where
        $s^{(d)}(t; \bm{\beta})$ is the limiting value of 
        $S^{(d)}(t; \bm{\beta})$, with $s^{(1)}(t; \bm{\beta})$ and $s^{(2)}(t; \bm{\beta})$  bounded, and $s^{(0)}(t; \bm{\beta})$ bounded away from $0$ for $t \in [0, \tau]$.
       \item[(C6).] 
       Positive-definiteness of the matrix $\bm{h}(\bm{\beta})=\int_0^{\tau} v(t; \bm{\beta})s^{(0)}(t; \bm{\beta})\lambda_0(t)dt$, for any $\bm{\beta}$ between $\bm{\beta}_0$ and $\widetilde{\bm{\beta}}$,       
       where 
     \begin{align*}
    v(t; \bm{\beta})= \left [\frac{s_k^{(2)}(\bm{\beta})}{s_k^{(0)}(\bm{\beta})}- \left \{\frac{s_k^{(1)}(\bm{\beta})}{s_k^{(0)}(\bm{\beta})}\right \}^{\otimes 2}\right ].   
     \end{align*}
       \end{itemize}
Conditions (C1)-(C6) are analogous to the regularity conditions employed in \citet{Andersen1982} 
and  \citet{Andersen1993} to prove the consistency and asymptotic normality of estimators from classical Cox models. 

\noindent
\textbf{Lemma 1}. \textit{If conditions (C1)-(C6) hold, then for each $\eta \geq 0$, $\hat{\bm{\beta}}(\eta)-\bm{\beta}^{*}(\eta)=O_p(n^{-1/2})$.}

Lemma 1 is a standard results showing that the CoxKL estimator $\hat{\bm{\beta}}(\eta)$ is consistent for the unique maximum for the expectation of the CoxKL objective function $\bm{\beta}^{*}(\eta)$. By Proposition 1, the CoxKL objective function \eqref{eq:obj_eta} retains a similar form as that of a classical Cox model, thus, Lemma 1 can be easily verified following standard arguments on the asymptotic results of maximum partial-likelihood estimator for classical Cox model under standard regularity conditions \citep{Andersen1993}.

\noindent
\textbf{Lemma 2}. \textit{If conditions (C1)-(C6) hold, and if $\eta = O(n^{-1/2})$, then as $n \rightarrow \infty$,
\begin{equation}
    n^{1/2} \{\hat{\bm{\beta}}(\eta)-\bm{\beta}^{*}(\eta) \} \overset{d}{\rightarrow} \mathcal{N}(\bm{0}, \bm{G}),
\label{eq:lemma2-1}
\end{equation}
where $\bm{G}\equiv\lim_{n\rightarrow \infty}(1+\eta)^{-1}\bm{H}^{-1}(\bm{\beta}_0)\bm{h}(\bm{\beta}_0)\bm{H}^{-1}(\bm{\beta}_0)$. If $\eta = o(n^{-1/2})$, then as $n \rightarrow \infty$,
\begin{equation}
    n^{1/2}\{\hat{\bm{\beta}}(\eta)-\bm{\beta}_0\}\overset{d}{\rightarrow} \mathcal{N}(\bm{0}, \bm{G}).
\label{eq:lemma2-2}
\end{equation}
}

Lemma 2 shows the asymptotic distribution of $\hat{\bm{\beta}}(\eta)$. When 
$\bm{\Delta}=\bm{0}$, $\hat{\bm{\beta}}(\eta)$ is an unbiased estimator for $\bm{\beta}_0$. In contrast, when $\bm{\Delta} \neq \bm{0}$ under population heterogeneity, as $n \rightarrow \infty$ and with the order of $\eta = o(n^{-1/2})$, $\hat{\bm{\beta}}(\eta) \overset{p}{\rightarrow} \bm{\beta}_0$. Moreover, the CoxKL estimator $\hat{\bm{\beta}}(\eta)$ yields a considerable efficiency gain over $\hat{\bm{\beta}}$ by incorporating external risk information, which can be easily seen from the form of asymptotic variance of $\hat{\bm{\beta}}(\eta)$. We then have the following Theorem 1, which states the advantage on the estimation of the CoxKL method.

\noindent
\textbf{Theorem 1}. \textit{If conditions (C1)-(C6) hold, then there exists a range of $\eta >0$, on which the asymptotic mean squared error (aMSE) of $\hat{\bm{\beta}}(\eta)$ is strictly less than that of $\hat{\bm{\beta}}$. Moreover, when $\bm{\Delta}=\bm{0}$, the aMSE of $\hat{\bm{\beta}}(\eta)$ is monotonically decreasing in $[0,\infty)$; when $\bm{\Delta}\neq\bm{0}$, the aMSE of $\hat{\bm{\beta}}(\eta)$ is monotonically decreasing in $[0,\eta^*)$ and monotonically increasing in $[\eta^*, \infty)$, with
\begin{equation*}
    \eta^*=\frac{\text{trace}\{\bm{h}^{-1}(\bm{\beta}_0)\}}{2n\bm{\Delta}^\top\bm{h}(\bm{c})^{\top}\bm{h}^{-2}(\bm{\beta}_0)\bm{h}(\bm{c})\bm{\Delta}-\text{trace}\{\bm{h}^{-1}(\bm{\beta}_0)\}},
\end{equation*}
where $\bm{c}\in \mathbb{R}^p$ is a vector between $\bm{\beta}_0$ and $\tilde{\bm{\beta}}$ such that 
\begin{equation*}
    \frac{1}{n} \sum_{k=1}^K \left \{\frac{S^{(1)}(t_k; \widetilde{\bm{\beta}})}{S^{(0)}(t_k; \widetilde{\bm{\beta}})} - \frac{S^{(1)}(t_k; \bm{\beta}_0)}{S^{(0)}(t_k; \bm{\beta}_0)}\right \}= - \bm{H}(\bm{c})\bm{\Delta}.
\end{equation*}
}

Theorem 1 implies that the CoxKL method improves the aMSE given an appropriate range of $\eta$ regardless of the potential heterogeneity. When $\bm{\Delta}=0$, $\hat{\bm{\beta}}(\eta)$ is unbiased, and it's intuitive that $\text{aMSE}(\hat{\bm{\beta}}(\eta)) <\text{aMSE}(\hat{\bm{\beta}})$ for all $\eta>0$ since the CoxKL method improves the estimation efficiency as proved in Lemma 2. Moreover, even if the external and internal information are heterogeneous and deviate  from each other, there exists a range of $\eta>0$ such that the CoxKL estimator $\hat{\bm{\beta}}(\eta)$ achieves a smaller aMSE for the model regression parameters compared to the maximum partial-likelihood estimator $\hat{\bm{\beta}}$ using the internal data only. Furthermore, as mentioned in Section~\ref{sub:external}, we do not require that the internal data and the external risk score share the same set of covariates. Since $\widetilde{\bm{\beta}}$ can be a sparse vector (i.e. some entries of $\widetilde{\bm{\beta}}$ are zero), we allow internal data to include additional predictors.


\section{Simulation}
\label{s:simu}
To evaluate the performance of the proposed CoxKL method, we conducted two sets of simulation studies. In Setting I, we focused on the evaluation of the estimation performance of CoxKL, while in Setting II, we evaluated the prediction performance of CoxKL under different scenarios, including the integration with various types of external risk score models (CoxPH and XGBoost \citep{Chen2016}) and multiple external risk scores. The tuning parameter $\eta$ was selected by the 5-fold cross validation with the V\&VH cross-validated partial likelihood as the measure of model performance.  In addition to the internal and the external models, we compared the performance of the CoxKL method with the stacked regression method \citep{debray14}, which includes the prediction of the external model as a covariate in the integration model. We evaluated the model performance on an independent testing dataset, which was simulated from the same distribution as the internal data. The simulation studies were replicated 500 times.

\subsection{Setting I: Evaluation of the estimation performance}
\label{sub:simu_est}

In the first set of simulations, we considered 6 covariates $\bm{Z}_1, \cdots, \bm{Z}_6$, where $\bm{Z}_1$, $\bm{Z}_2$ were continuous variables simulated from a multivariate normal distribution with zero mean, unit variance and a first-order autoregressive (AR1) correlation structure with the auto-correlation parameter 0.5. $\bm{Z}_{3}, \bm{Z}_{4}$ were binary variables with $Pr(\bm{Z}_i=1)=0.5$. Regarding $\bm{Z}_{5}$ and $\bm{Z}_{6}$, to simulate the situation where the internal cohort comes from a different underlying distribution from the external population, we denoted $\bm{Z}^{l}$ as a latent binary variable with $Pr(\bm{Z}^l_i=1)=p_l$, and then let $\bm{Z}_{5}$ and $\bm{Z}_{6}$ be two continuous variables simulated from normal distribution with unit variance and mean $2\bm{Z}^{l}$ and $-2\bm{Z}^{l}$, respectively. That is, the covariate distribution of the internal cohort could be different from that of the external cohort. Note that, as a latent variable, $\bm{Z}^{l}$ was unobserved in the simulated datasets. The survival time for the internal data was generated from a Cox model with a hazard function of 
\begin{eqnarray}
\lambda(t|\bm{Z}_{1i},\dots, \bm{Z}_{6i})=2t\times\exp(\beta_1\bm{Z}_{1i}+\beta_2\bm{Z}_{2i}+\beta_3\bm{Z}_{3i}+\beta_4\bm{Z}_{4i}+\beta_5\bm{Z}_{5i}+\beta_6\bm{Z}_{6i}),
\label{eq:sim_set1}
\end{eqnarray}
where $\bm{\beta}^{I}=(\beta_1, \beta_2, \beta_3, \beta_4, \beta_5, \beta_6)^\top=(0.3, -0.3, 0.3, -0.3, 0.3, -0.3)^\top$. The censoring times were generated from a uniform distribution with varying upper bounds to control for different censoring proportions. We set $p_l^{I}=1$ for the internal cohort. The true generating mechanism for the external cohort was the same as \eqref{eq:sim_set1}, but the external model was determined from a large dataset with varying $p_l^E$ (\% homogeneous) and $\bm{Z}^{E}$ (covariates used in the external model). Specifically, we considered the following external settings:
\begin{itemize}
        \item[(E1).] The covariate distribution of the external population is the same as the internal cohort, that is, $p_l^E=1$; and the external model is the true model: $\bm{Z}^{E}=   (\bm{Z}_{1},\bm{Z}_{2},\bm{Z}_{3},\bm{Z}_{4},\bm{Z}_{5},\bm{Z}_{6})^\top$.
        \item[(E2).] The covariate distribution of the external population is slightly different from the internal cohort with $p_l^E=0.5$; and the external model is a misspecified model: $\bm{Z}^{E}=(\bm{Z}_1, \bm{Z}_3, \bm{Z}_5, \bm{Z}_6)^\top$.
        \item[(E3).] The covariate distribution of the external population is completely different from the internal cohort with $p_l^E=0$; and the external model is a misspecified model: $\bm{Z}^{E}=(\bm{Z}_1, \bm{Z}_5)^\top$. 
\end{itemize}


\begin{table}[htbp]
\caption{Comparison of estimation performance under simulation setting I. Internal = Cox proportional hazards model based on the internal data only; CoxKL = integrated model by the proposed CoxKL method. For the internal data, the sample size varied from $\{50, 100\}$, and the censoring rate varied from $\{30\%, 60\%\}$. For external model, setting (E1) corresponds to the good-quality external model, setting (E2) corresponds to the fair-quality external model, while setting (E3) corresponds to the poor-quality external model. The simulation study was replicated 500 times. Bias = average empirical bias; SE = average empirical standard error; MSE = average empirical mean squared error. C-Index = Harrel's C-Index. Lower Bias, SE, and MSE indicated better estimation performance. Higher C-index indicated better prediction performance.\label{tab:est_true}}
\setlength{\tabcolsep}{5pt} 
\renewcommand{\arraystretch}{1.2} 
\vspace*{0.5cm}
\centering
\begin{tabular}{rccccccc}
\hline
 \hline 
  & \multicolumn{3}{c}{External model} & & & & \\ 
    \cline{2-4}    Method & Setting & \% Homogeneous & Covariates  & Bias & SE & MSE & C-Index \\ 
\hline 
\hline 
\rowcolor{Gray2}
\multicolumn{8}{l}{Internal data: sample size = 50, censoring rate = 60\%}\\
  \hline
 Internal& --- & --- & --- &0.057 & 0.475 & 0.229 & 0.594 \\ \cline{1-8}
 CoxKL& (E1) & 100 &$(\bm{Z}_1, \bm{Z}_2,\bm{Z}_3,\bm{Z}_4,\bm{Z}_5,\bm{Z}_6)$ &0.023 & 0.136 & 0.019 & 0.640 \\ 
 & (E2) & 50& $(\bm{Z}_1, \bm{Z}_3,\bm{Z}_5,\bm{Z}_6)$ & 0.101 & 0.165 & 0.038 & 0.624 \\ 
 & (E3) & 0&$(\bm{Z}_1, \bm{Z}_5)$ & 0.145 & 0.192 & 0.058 & 0.605 \\ \hline \rowcolor{Gray2}
  \multicolumn{8}{l}{Internal data: sample size = 50, censoring rate = 30\%}\\
  \hline
  Internal & ---& ---& ---&0.040 & 0.354 & 0.127 & 0.608 \\ \cline{1-8}
  CoxKL& (E1) & 100& $(\bm{Z}_1, \bm{Z}_2,\bm{Z}_3,\bm{Z}_4,\bm{Z}_5,\bm{Z}_6)$ & 0.022 & 0.114 & 0.014 & 0.642 \\ 
  & (E2) & 50 & $(\bm{Z}_1, \bm{Z}_3,\bm{Z}_5,\bm{Z}_6)$ & 0.088 & 0.149 & 0.030 & 0.627 \\ 
  & (E3) & 0 & $(\bm{Z}_1, \bm{Z}_5)$ & 0.129 & 0.176 & 0.048 & 0.612 \\ \hline \rowcolor{Gray2}
  \multicolumn{8}{l}{Internal data: sample size = 100, censoring rate = 30\%}\\  
    \hline
  Internal &--- &--- &--- & 0.019 & 0.207 & 0.043 & 0.628 \\ \cline{1-8}
  CoxKL & (E1) & 100 &$(\bm{Z}_1, \bm{Z}_2,\bm{Z}_3,\bm{Z}_4,\bm{Z}_5,\bm{Z}_6)$ & 0.023 & 0.066 & 0.005 & 0.645 \\
    & (E2) & 50 &$(\bm{Z}_1, \bm{Z}_3,\bm{Z}_5,\bm{Z}_6)$ & 0.079 & 0.111 & 0.019 & 0.633 \\ 
    & (E3) & 0 &$(\bm{Z}_1, \bm{Z}_5)$ & 0.093 & 0.138 & 0.028 & 0.627 \\ \hline \hline 
\end{tabular}
\end{table}

\begin{figure}[htbp]
\begin{center}
\centerline{\includegraphics[width=13cm]{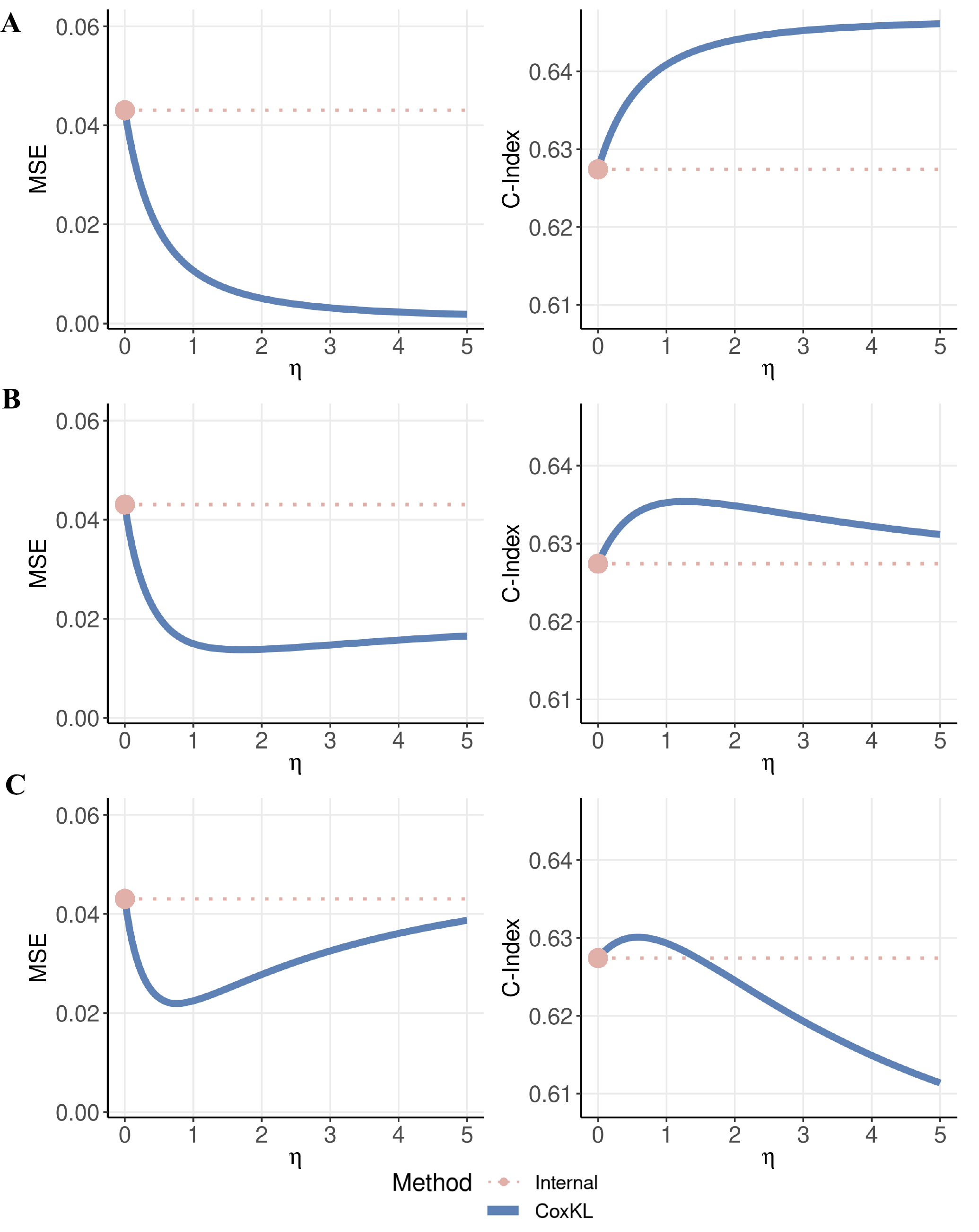}}
\end{center}
\caption{Comparison of MSE and C-index with different $\eta$ values under simulation setting I. Internal = Cox proportional hazards model based on the internal data only; CoxKL = integrated model by the proposed CoxKL method. 
Panel A corresponds to external Setting (E1) with one good-quality external model. Panel B corresponds to external Setting (E2) with one fair-quality external model. Panel C corresponds to external Setting (E3) with one poor-quality external model. Details of external settings can be found in Section~\ref{sub:simu_est}. The simulation study was replicated 500 times. MSE = average empirical mean squared error. C-Index = Harrel's C-Index. Lower MSE indicated better estimation performance. Higher C-index indicated better prediction performance.\label{fig:kl_simu_4_true}}
\end{figure}
Table \ref{tab:est_true} and Figure \ref{fig:kl_simu_4_true} summarize the results of simulation Setting I, including average empirical biases (Bias), average empirical standard errors (SE), and average empirical mean squared errors (MSE) for the estimated regression coefficients $\bm{\beta}^{I}=(\beta_1, \beta_2, \beta_3, \beta_4, \beta_5, \beta_6)^\top$. In addition, we also presented the Harrel's C-Index \citep{Harrel1996}. As shown in Table \ref{tab:est_true}, under all three external settings regarding good, fair, and poor external model quality, the integration model by the CoxKL method attained
better estimation efficiency than the internal model, with smaller SE and MSE; moreover, the SE and MSE of the CoxKL method decreased as the quality of the external risk information increased. Incorporating external information from homogeneous models (external setting E1) reduced the bias when the quality of the internal cohort was limited. Unsurprisingly, when the external models and internal cohort were heterogeneous, incorporating information from misspecified external risk scores increased the bias. However, due to the bias and variance trade-off, the MSE of the CoxKL estimator still decreased. Compared to the internal model, the CoxKL method also improved the prediction performance by incorporating external information, with a higher C-Index. As the internal sample size increased or the internal censoring rate decreased, the gain on estimation efficiency and prediction accuracy by incorporating external information via the CoxKL method decreased, as expected. As shown in Figure \ref{fig:kl_simu_4_true}, there existed a range of integration weight $\eta$ that the integration model by CoxKL method reduced the MSE and increased the C-Index compared to the internal model. With a decrease in the external information quality, the range of $\eta$ that improved the model performance shrank; that is, the optimal integration weight of the external risk scores decreased. However, the CoxKL method outperformed the internal model in all cases with a properly selected integration weight $\eta$.

\subsection{Setting II: Evaluation of the prediction performance}
\label{sub:simu_pred}

In Setting II, we included 5 covariates $\bm{Z}_{1}, \dots,\bm{Z}_{5}$, which were simulated from the same distributions as in Setting I. In addition to the main effect, we also considered the interaction term $\bm{Z}_{1}*\bm{Z}_{5}$ in the data generating model. Since in practice, we usually cannot observe and collect all the variables in the true model, here we simulated two additional variables, $\bm{Z}^{u1}$ and $\bm{Z}^{u2}$,
which contributed to the true underlying generating distribution of the data, but were not \textit{observed} and \textit{collected} in the simulated datasets. $\bm{Z}^{u1}$ was a binary variable with $Pr(\bm{Z}^{u1}_i=1)=0.5$, and $\bm{Z}^{u2}$ came from a standard normal distribution. 
The true generating model for the internal data was a Cox model with a hazard function of 
\begin{eqnarray}
\nonumber \lambda(t|\bm{Z}_{1i},\dots, \bm{Z}_{5i}, \bm{Z}^{u1}_i, \bm{Z}^{u2}_i)=2t&\times&{}\exp(\beta_1\bm{Z}_{1i}+\beta_2\bm{Z}_{2i}+\beta_3\bm{Z}_{3i}+\beta_4\bm{Z}_{4i}+\beta_5\bm{Z}_{5i}\\ 
&&{}+\beta_6\bm{Z}_{1i}*\bm{Z}_{5i}+\beta_{u1}\bm{Z}^{u1}_{i}+\beta_{u2}\bm{Z}^{u2}_{i}),
\label{eq:sim_set2}
\end{eqnarray}
where $(\beta_1, \beta_2, \beta_3, \beta_4, \beta_5, \beta_6, \beta_{u1}, \beta_{u2})^\top=(0.3, -0.3, 0.3, -0.3, -0.3, 0.5, 1, 1)^\top$. The true generating mechanism for the external cohort was the same as \eqref{eq:sim_set2}. However, the external models were determined from large datasets with varying $p_l$ and $\bm{Z}^{E}$. 

\begin{figure}[htbp]
\begin{center}
\centerline{\includegraphics[width=13cm]{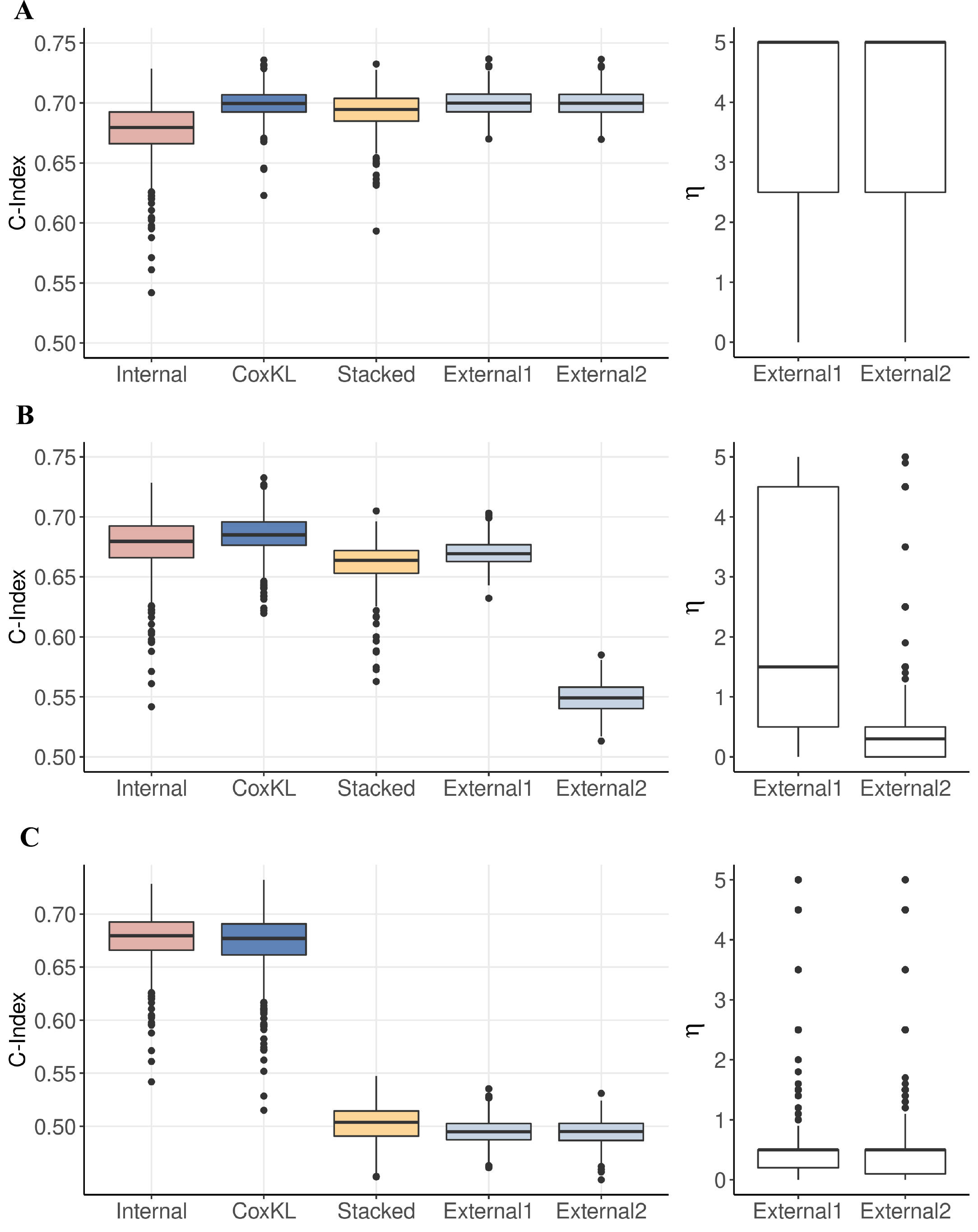}}
\end{center}
\caption{Comparison of prediction performance under simulation setting II. Panel A corresponds to External setting (E4) with two good-quality external models. Panel B corresponds to External setting (E5) with one good-quality external model and one poor-quality external model. Panel C corresponds to External setting (E6) with two poor-quality external models. Details of external settings can be found in Section~\ref{sub:simu_pred}. Internal = Cox proportional hazards model based on the internal data only; External1 was derived from a XGBoost model, and External2 was derived from a Cox model, as specified in settings (E4), (E5), and (E6); CoxKL = integrated model by CoxKL method incorporates both External1 and External2; Stacked = stacked model by stacking regression incorporates both External1 and External2. Plots on the left panel present prediction performance of each method. Higher C-index indicated better prediction performance. Plots on the right panel present the optimal selection of the integration weight $\eta$ by CoxKL. Larger $\eta$ indicated the external model contributed more information to the integrated model. Simulation was replicated 500 times.
\label{fig:kl_simu_1_1}}
\end{figure}

We first considered the settings with two external risk scores derived from different types of models. Specifically, in each setting, the External1 was derived from a XGBoost model, and the External2 was derived from a CoxPH model. Two extreme settings with extremely well or poorly performing external models were explored. Details are given as follows:
\begin{itemize}
        \item[(E4).] Two good-quality external models: External1: $p_l^E=1$, and $\bm{Z}^{E}=(\bm{Z}_{1},\bm{Z}_{2},\bm{Z}_{3},\bm{Z}_{4},\bm{Z}_{5})^\top$; External2: $p_l^E=1$, and $\bm{Z}^{E}=(\bm{Z}_{1},\bm{Z}_{2},\bm{Z}_{3},\bm{Z}_{4},\bm{Z}_{5})^\top$.
        \item[(E5).] One good-quality external model and one poor-quality external model. External1: $p_l^E=0.25$, and $\bm{Z}^{E}=(\bm{Z}_{1},\bm{Z}_{3},\bm{Z}_{5})^\top$; External2: $p_l^E=0.25$, and  $\bm{Z}^{E}=(\bm{Z}_{2},\bm{Z}_{4})^\top$.
        \item[(E6).] Two poor-quality external models: External1: $p_l^E=0$, and $\bm{Z}^{E}=(\bm{Z}_{2},\bm{Z}_{4})^\top$; External2: Null model.
\end{itemize}

For the internal cohort, we set the sample size as 50, the censoring rate as around 30\%, and $p_l^I=1$. As shown in Figure \ref{fig:kl_simu_1_1}, under extreme scenarios such as both external risk scores were derived from extremely well or poorly performing models, the CoxKL method could achieve comparable prediction performance as the external or internal model by selecting an extremely large or small integration weight $\eta$. Moreover, when integrating two external risk scores with different prediction performances, the CoxKL method could optimally determine the weight of information integration, assigning more weight to the better performing external risk score, and resulting in improved prediction performance with the highest C-Index. Incorporating information from external models resulted in a narrower interquartile range (IQR) compared with that of the internal model, which indicates that the proposed method reduced the variance of prediction. Note that under all three settings, the External1 risk scores were derived by the XGBoost, which indicates that the CoxKL method worked well when integrating a risk score derived from a machine learning model.

\begin{figure}[htbp]
\begin{center}
\centerline{\includegraphics[width=16cm]{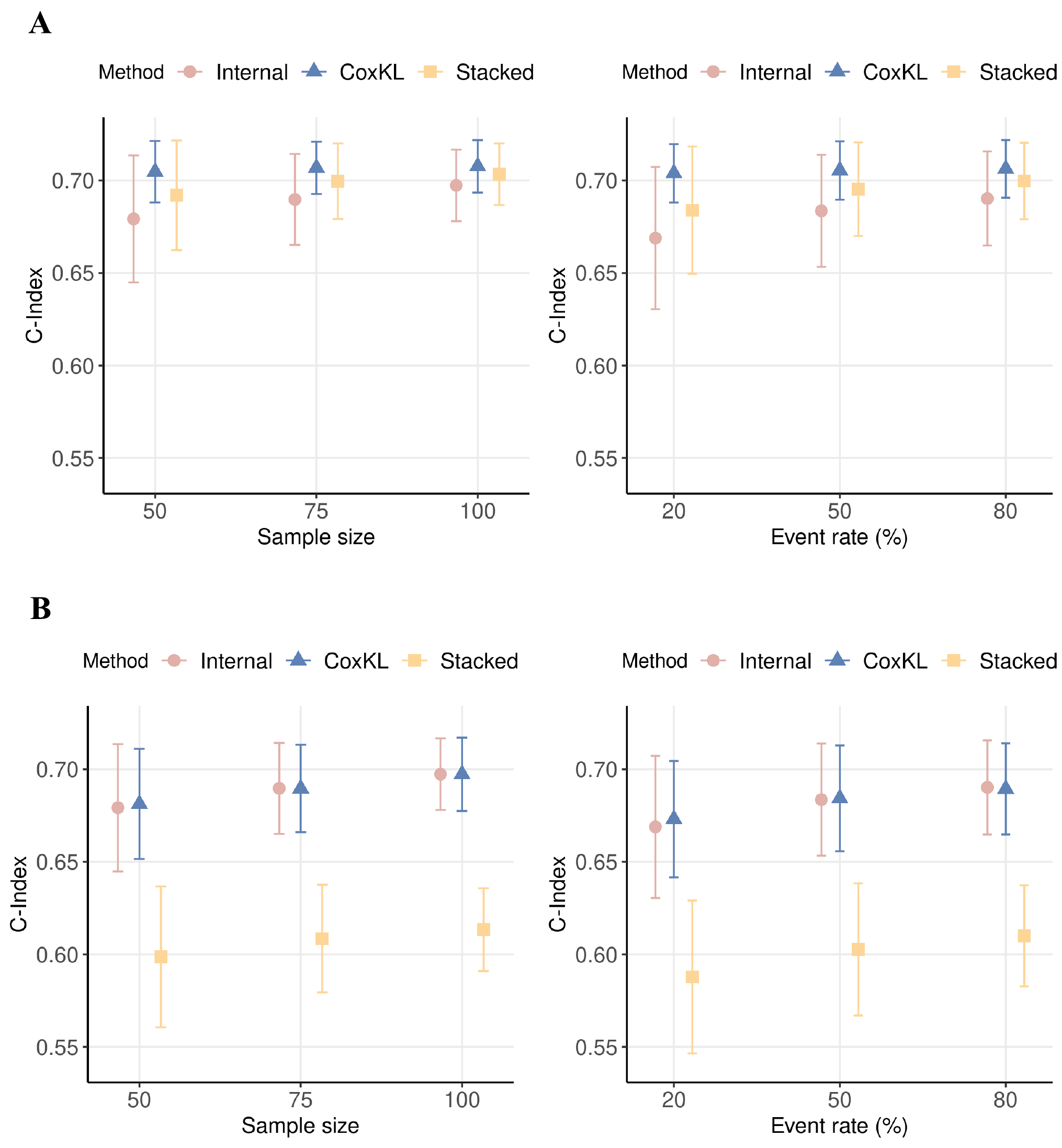}}
\end{center}
\caption{Comparison of prediction performance under simulation setting II. For plots on the left panel, the sample size of the internal data varied from $\{50, 75, 100\}$ with the event rate (1 $-$ censoring rate) set to be 40\%. For plots on the right panel, the event rate of the internal data varied from $\{40\%, 60\%, 80\%\}$ with the sample size set to be 50. Panel A corresponds to a fair-quality external model setting with $p_l^{E} = 1$ ($\% \text{Homogeneous} = 100$) and covariates as $\bm{Z}^{E}=(\bm{Z}_1, \bm{Z}_3, \bm{Z}_5)$. Panel B corresponds to a poor-quality external model setting with $p_l^{E} = 0$ ($\% \text{Homogeneous} = 0$) and covariates as $\bm{Z}^{E}=(\bm{Z}_1, \bm{Z}_3, \bm{Z}_5)$. Higher C-index indicates better prediction performance. Results were averaged for 500 replicates. 
\label{fig:kl_simu_2}}
\end{figure}

Then we considered different settings in terms of sample sizes and censoring rates of the internal cohort. We set $p_l^I=1$ for the internal cohort, and let $p_l^E$ varying from $\{0, 1\}$ for the external cohort. The external covariates were set as $\bm{Z}^{E}=\{\bm{Z}_1, \bm{Z}_2, \bm{Z}_3\}$. The sample size of the internal cohort varied from $\{50, 75, 100\}$, and the censoring rate of the internal data varied from $\{20\%, 50\%, 80\%\}$. The simulation results are shown in Figure \ref{fig:kl_simu_2}. With increasing censoring rates of the internal data, the CoxKL method outperformed the internal and stacked models with higher C-index and smaller variance of prediction. The prediction performances of the internal and stacked models decreased rapidly as the censoring rate increased or the sample size decreased. In addition, for fairly small datasets, incorporating information from the external prediction model by the CoxKL method improved the prediction performance over the internal model. As expected, the improvement decreased when the internal cohort contained more data.

\section{Application: Prostate Cancer Trans-ancestry Polygenic Hazard Score}
\label{s:real_data}

Prostate cancer is the most prevalent cancer and second leading cause of cancer death in American males \citep{Siegel2022}. Identifying high-risk population and offering personalized suggestions on when to initialize prostate cancer screening is important to reduce the cancer-related morbidity and mortality \citep{Schroder2014}. PHS46 has shown accurate discrimination of onset of prostate cancer in EUR males but reduced performance in non-EUR males, such as AFR males \citep{Huynh2021}. To improve the risk discrimination of prostate cancer in AFR males, we applied CoxKL to develop a trans-ancestry PHS by integrating PHS46 with a small-sized internal dataset of AFR individuals.

The PHS46 model is a Cox proportional hazards model developed using a genotype dataset consisting of EUR males. By analyzing associations between genetic factors and the age at diagnosis of prostate cancer, the model includes 46 SNPs and predicts prostate cancer risk in EUR males. Based on the external validation results, the PHS46 model performs poorly with a C-index of 0.55 in AFR males \citep{Karunamuni2021}. Thus, to improve the risk discrimination for prostate cancer in AFR males, an ancestry specific PHS model is desired.

The internal dataset obtained from the Michigan Genomics Initiative (MGI) included 1,036 AFR males with both genotype and phenotype data \citep{MGI}. The outcome was defined as the age at diagnosis of prostate cancer. For subjects who were not diagnosed with prostate cancer, the study follow-up was censored at the age of the last visit, and the overall censoring rate was $82.5\%$. Standard quality control was performed on the genotype data. Eventually, a total of 161,861
imputed SNPs passing a quality threshold of $R^2>0.3$, with a minor allele frequency $> 0.05$, and LD pruned with $R^2 > 0.1$, $+/- 500kb$ and $\text{step size}=1$ were considered in the analysis. In addition, we conducted a GWAS analysis to further identify the SNPs that may be associated with the outcome using univariate Cox regression models. Top 10,000 SNPs identified based on p-values from the GWAS results were included in the PHS model development.

With the moderate sample size, high censoring rate and high-dimensional covariate space, we considered incorporating PHS46 with the internal data by utilizing the proposed CoxKL-LASSO procedure (introduced in Section \ref{sub:Estimation}). The internal data was randomly split into training and testing sets with a ratio of 3:1. We developed PHS models on the training set and evaluated the model performance on the testing set. A total of 25 independent random splits were implemented. As comparisons, results of the PHS46 model and the internal model by applying a penalized Cox regression model with LASSO on the internal data are also presented.

\begin{figure}[htbp]
\begin{center}
\centerline{\includegraphics[width=16cm]{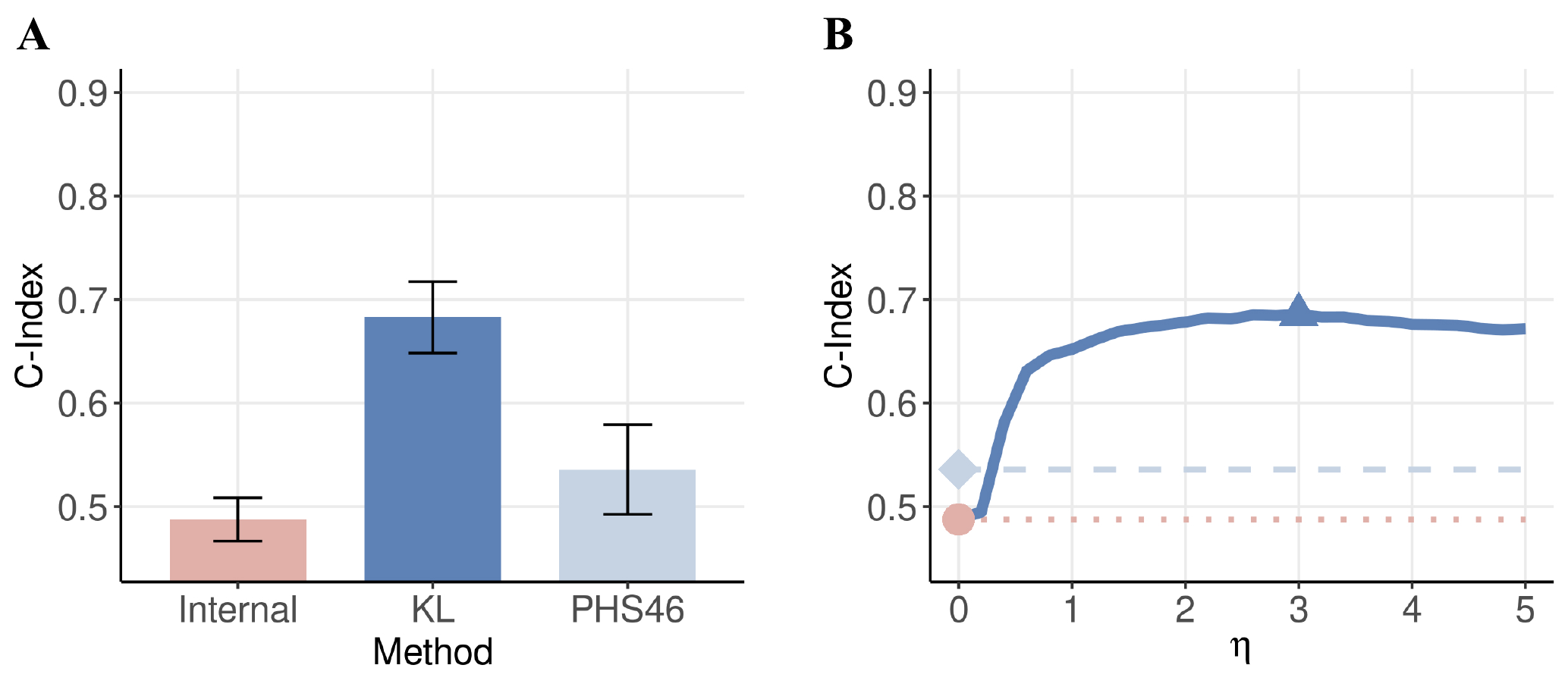}}
\end{center}
\caption{Results of the PHS models predicting the onset of prostate cancer in African American ancestry individuals. Sample size of the internal data is $n=1036$, and the censoring rate is $82.5\%$. Panel A = comparison of prediction performance across internal, CoxKL, and PHS46 models. Panel B = performance of the integrated model by CoxKL-LASSO method with different weight $\eta$. 
CoxKL = integrated model by the proposed CoxKL-LASSO incorporating PHS46 with the internal data. Internal = Cox LASSO model fitted on the internal data only. PHS46 = the published PHS model for prostate cancer in European ancestry. A total of 10,000 SNPs after standard quality control and screening were included in the developments of the PHS models. The internal data was randomly split into training and testing sets with a ratio of 3:1. We developed PHS models on the training set and evaluated the model performance on the testing set. A total of 25 independent random splits were implemented. The optimal $\eta$ was selected to be around $3$. Larger C-index indicates better prediction performance.
\label{fig:real_data1}}
\end{figure}

\begin{figure}[htbp]
\begin{center}
\centerline{\includegraphics[width=8cm]{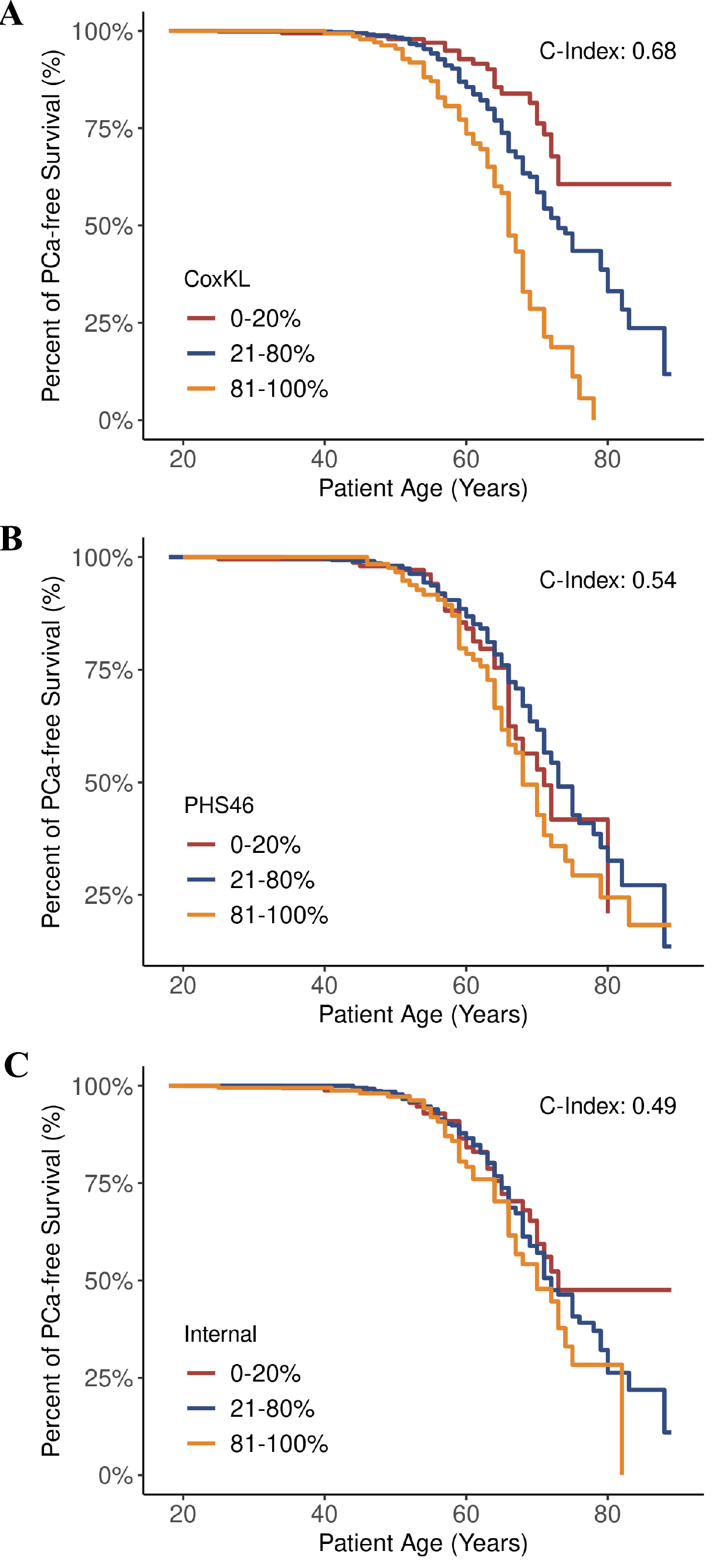}}
\end{center}
\caption{Risk discrimination of the PHS models on the onset of prostate cancer in African American ancestry males. Kaplan-Meier patient Pca-free survival curves by each PHS model are presented. Panel A = risk discrimination by the proposed CoxKL-LASSO method. Panel B = risk discrimination by the PHS46 model. Panel C = risk discrimination by the internal model. PHS46 = the published PHS for prostate cancer in European ancestry. CoxKL = the integrated model by the proposed CoxKL-LASSO incorporating PHS46 with the internal data. Internal = Cox LASSO model fitted on the internal data only. PCa = prostate cancer. High-risk populations (subjects with PHS score ranging from $81-100\%$; represented by the orange line) were clearly identified by the proposed CoxKL method. Larger C-index indicates better prediction performance. \label{fig:real_data2}}
\end{figure}

As shown in Figure \ref{fig:real_data1}, the proposed method had the best prediction performance compared to the internal model and PHS46 model. Specifically, when the dimension of the covariate space was much larger than the sample size, it was difficult to develop a proper PHS model based on the internal small-sized data only. Therefore, the CoxKL method incorporated more information from PHS46 with an optimal $\eta$ at around $3$. Furthermore, as shown in Figure \ref{fig:real_data2}, the CoxKL method also achieved better risk discrimination of prostate cancer in AFR males than both the internal and PHS46 models. High-risk populations (subjects with PHS score ranging from $81-100\%$) were clearly identified by the proposed CoxKL method,
which indicates that our method can be used for better personalized suggestions on when to initialize prostate cancer screening for AFR males. 

\section{Discussion}
\label{s:dis}

In this paper, we proposed the CoxKL method to incorporate external risk scores derived from published prediction models with internal time-to-event data. Partial-likelihood-based KL information is utilized to provide an overall assessment of heterogeneity across populations, and a penalized objective function is developed to balance the contribution of different information sources. 
Extensive simulation studies 
show the advantages of the CoxKL method, including improvements on estimation efficiency and prediction accuracy. 

The CoxKL method is computationally efficient and, thus, can be easily extended to integration problems under high-dimensional settings. 
We presented CoxKL-LASSO for developing trans-ancestry PHS prediction models in non-EUR populations. We applied our method to develop a trans-ancestry PHS model for prostate cancer by integrating PHS46, a previously published EUR-based PHS, with a small-sized AFR genotype data. The prostate cancer risk discrimination in AFR males has been substantially improved (C-Index from 0.49 [Internal] and 0.54 [PHS46] to 0.68 [CoxKL]). The proposed AFR-based trans-ancestry PHS can clearly distinguish subjects with high prostate cancer risk.

As the CoxKL method summarizes external risk information into personalized risk scores, it provides a unified framework for integrating risk information from various types of external models, including prediction models by machine learning algorithms, as long as the external models can provide personalized risk scores for each subject in the internal data. Moreover, as shown by both theoretical properties and simulation studies, the CoxKL method allows internal data include additional covariates than that used in external risk score models. The CoxKL method can also be easily extended to the integration with risk information from multiple external models, details are given in Supporting Information. 

In summary, the proposed CoxKL method serves as a general and flexible integration tool for time-to-event data, with an important application in developing trans-ancestry PHS models and enhancing risk discrimination in non-EUR populations.

\backmatter


\section*{Acknowledgements}
The authors acknowledge the Michigan Genomics Initiative participants, Precision Health at the University of Michigan, the University of Michigan Medical School Central Biorepository, 
the University of Michigan Medical School Data Office for Clinical and Translational Research, the University of Michigan Advanced Genomics Core for providing data storage and specimen storage, management, processing, and distribution services, and the Center for Statistical Genetics in the Department of Biostatistics at the School of Public Health for genotype data curation, imputation, and management in support of the research reported in this publication. This work was partially supported by the National Institute of Diabetes and Digestive and Kidney Diseases under Grant R01DK129539.
\bibliographystyle{biom} 
\bibliography{CoxKL.bib}



\label{lastpage}

\end{document}